\documentclass[12pt]{article}
\usepackage{epsfig}
\usepackage{cite}
\usepackage{amssymb}
\usepackage{wrapfig}
\usepackage{graphics}
\def\g{\gamma}
\def\a{\Gamma}

\def\pp{\pi^+}
\def\pn{\pi^-}
\def\po{\pi^0}
\def\gp{\g d\to\pp\g nn}
\def\gn{\g d\to\pn\g pp}
\def\go{\g d\to\po\g pn}
\def\beq{\begin{eqnarray}}
\def\eeq{\end{eqnarray}}

\begin{document}
\begin{center}
{\bf SEARCH FOR NARROW SIX-QUARK STATES IN THE REACTIONS   
${\bf \g d\to \pi\g NN}$} \\
\underline{L.V.~Fil'kov},$^{1}$ V.L. Kashevarov,$^{1,2}$ M. Ostrick $^{2}$ \\
\end{center}
\noindent
 $^1$--{\it Lebedev Physical Institute, Moscow, Russia} \\
 $^2$--{\it Institut f\"ur Kernphysik, University of Mainz, Germany} \\ 

\begin{center} 
{\bf Abstract}
\end{center}
We study the reactions $\g d\to\pi\g NN$ with the aim
to search for six-quark states, the decay of which into two nucleons is
forbidden by the Pauli exclusion principle. These states with the masses
$M<2m_N+m_{\pi}$ should mainly decay by emitting a photon. This is a new class
of metastable six-quark states with the decay widths $<< 1$keV. 
The recent experiment at Proton Linear Accelerator of INR (Moscow)  
suggests a possibility of the existence of such states
with masses 1904, 1926, and 1942 MeV.
The proposed experiment at
MAMI-C could provide an unique opportunity to observe such dibaryon states
in mass region up to 2000 MeV and determine their masses and quantum
numbers. 

\section{Introduction}
\label{sec:intro}

The possibility of the existence of multiquark states was predicted by QCD
inspired models \cite{jaf,muld}. These works initiated a lot of experimental
searches for six-quark states (dibaryons). Usually authors of these works 
looked for
dibaryons in the $NN$ channel (see for review ref. \cite{tat1}).
Such dibaryons have decay widths from a few up
to hundred MeV. Their relative contributions are small enough and the
background contribution is big and uncertain as a rule. It often
leads to contradictory results.

We consider six-quark states,
a decay of which into two nucleons is forbidden by the Pauli exclusion
principle.
Such states satisfy the following condition:
\begin{equation}
(-1)^{T+S}P=+1
\end{equation}
where $T$ is the isospin, $S$ is the spin of the nucleon pair
inside a six-quark state, and $P$ is the
dibaryon parity. In the $NN$ channel, these six-quark states would
correspond to the following forbidden states:
even singlets and odd triplets with the isotopic spin $T=0$ as well as
odd singlets and even triplets with \mbox{$T=1$}.
Such six-quark states with the masses \mbox{$M < 2m_{N}+m_{\pi}$}
($m_N (m_{\pi}$) is the nucleon (pion) mass) could mainly decay
by emission of a photon. This is a new class of metastable six-quark states
with the decay widths \mbox{$<< 1$keV} \cite{fil1,fil2}. 
Such states were called "supernarrow dibaryons" (SND).

The experimental discovery of the SNDs would have important consequences
for particle and nuclear physics and astrophysics. This would lead to
a deeper understanding of the evolution of compact stars and the new
possibility of quark-gluon plasma production in their interior. 
In nuclear physics there would be a new concept: SND-nuclei.

In the framework of the MIT bag model, Mulders et al. \cite{muld} calculated
the masses of different dibaryons, in particular, $NN$-decoupled dibaryons.
They predicted dibaryons $D(T=0;J^P =0^{-},1^{-},2^{-};M=2.11$ GeV) \ and
$D(1;1^{-};2.2$ GeV) corresponding to the forbidden states $^{13}P_J$
and $^{31}P_1$ in the $NN$ channel. However, the dibaryon masses obtained 
by them exceed the pion production threshold.
Therefore, these dibaryons preferentially decay into the $\pi NN$ channel
and their decay widths are larger than 1 MeV.

Using the chiral soliton model, Kopeliovich \cite{kop} predicted that
the masses of \mbox{$D(1,1^+)$} and $D(0,2^+)$ SNDs could
exceeded the two nucleon mass by 60 and 90 MeV, respectively.
These values are lower than the pion production threshold.

In the framework of the canonically quantized biskyrmion model,
Krupnovnickas {\em et al.} \cite{riska} obtained an indication on
possibility
of the existence of one dibaryon with J=T=0 and two dibaryons with J=T=1
with masses smaller than $2m_N+m_{\pi}$.

Unfortunately, all predicted values of the dibaryon masses are strongly model
dependent. Therefore, only an experiment could answer the question about
the existence of SNDs and determine their masses. 

For the first time, SNDs have been observed in the reactions
$pd\to p+pX_1$ and $pd\to p+dX_2$
\cite{konob,yad,prc,epj}.
The experiment was
carried out at the Proton Linear Accelerator of INR with 305 MeV
proton beam using the two-arm mass spectrometer TAMS. 

Several software cuts have been applied to the mass spectra in these works.
In particular, the authors limited themselves by the consideration of an
interval of the proton energy from the decay of the $pX_1$ states
and its narrow angle cone of the with respect to the direction of the 
dibaryon motion, which were determined 
by the kinematics of the SND decay into $\gamma NN$ channel.
Such cuts are very important as they provide a possibility to suppress the
contribution from the background reactions and random coincidences
essentially.

\begin{wrapfigure}{l}{0.5\textwidth}
\epsfxsize=0.48\textwidth       
\epsfysize=10cm                
\epsffile{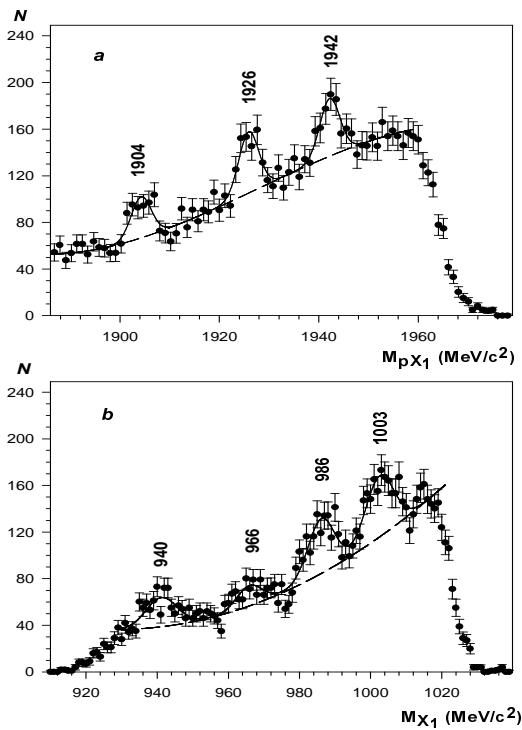}  
\caption{The missing mass $M_{pX_1}$ (a) and
$M_{X_1}$ (b) spectra of the reaction $pd\to p+pX_1$.
The dashed and solid curves are results of interpolation by polynomials
(for the background) and Gaussian (for the peaks), respectively.}
\label{fig1}
\end{wrapfigure} 
As a result,
three narrow peaks in missing mass spectra were observed (Fig. 1a) at
$M_{pX_1}=1904\pm 2$, $1926\pm 2$, and $1942\pm 2$ MeV  with widths equal to
the experimental
resolution ($\sim 5$ MeV) and with numbers of standard deviations (SD) of
6.0, 7.0, and 6.3, respectively.
It should be noted that the dibaryon peaks at $M=1904$ and 1926 MeV had
been observed earlier by same authors in ref. \cite{prc,konob,yad}
under different kinematical conditions. On the other hand, no noticeable
signal of the dibaryons were observed in the missing mass $M_{dX_2}$
spectra of the reaction $pd\to p+dX_2$.
The analysis of the angular distributions of the protons from the decay of
$pX_1$ states and the suppression observed of the SND decay into $\gamma d$
showed that the peaks found can be explained as a manifestation of 
isovector SNDs, the decay of which into two nucleons is forbidden
by the Pauli exclusion principle.

Additional information about the nature of the observed states
was obtained by studying the missing mass $M_{X_1}$ spectrum of the
reaction $pd\to p+pX_1$.
If the state found is a dibaryon decaying mainly into two nucleons then
$X_1$ is the neutron and the mass $M_{X_1}$ is equal to the neutron mass
$m_n$. If the value of $M_{X_1}$, obtained from the experiment, differs
essentially from $m_n$, then $X_1=\gamma+n$ and we have the additional
indication that the observed dibaryon is SND.

In the experimental missing mass $M_{X_1}$ spectrum besides the peak
at the neutron mass caused by the process $pd\to p+pn$,
a resonance-like behavior of the spectrum were observed at $966\pm 2$,
$986\pm 2$, and $1003\pm 2$ MeV \cite{epj}.
These values of $M_{X_1}$ coincide with
the ones obtained by the simulation. 
Hence, for all states under
study, we have $X_1=\gamma+n$ in support of the statement that the
dibaryons found are SNDs.

On the other hand, the peak at $M_{X_1}=1003\pm 2$ MeV corresponds to
the peak found in ref. \cite{tat2} and was attributed to an exotic
baryon state $N^*$ below the $\pi N$ threshold. In that work the authors
investigated the reaction $pp\to\pi^+pX$ and found
altogether three such states with masses 1004, 1044, and 1094 MeV.
Therefore, if the exotic baryons with anomalously small masses really
exist, the observed peaks at 966, 986, and 1003 MeV might be a manifestation
of such states.
The existence of such exotic states, if established,
would fundamentally change our understanding of the
quark structure of hadrons \cite{bald,walch}.
However, the experiments on single nucleons have
not observed any significant structure \cite{lvov,jiang,kohl}. Therefore,
the question about a nature of peaks observed in \cite{epj,tat2} is
open still.

In ref. \cite{khr} dibaryons with exotic quantum numbers were searched for
in the process $pp\to pp\gamma\gamma$. The experiment was performed with
a proton beam from the JINR Phasotron at an energy of about 216 MeV. The
energy spectrum of the photons emitted at $90^{\circ}$ was measured.
As a result, two peaks were observed in this spectrum. This behavior
of the photon energy spectrum was interpreted as a signature of the exotic
dibaryon resonance $d_1$ with a mass of about 1956 MeV and possible isospin
$T=2$. 

On the other hand, an analysis \cite{cal} of the Uppsala proton-proton
bremsstrahlung data looking for the presence of a dibaryon in the mass range
from 1900 to 1960 MeV only gave upper limits of 10 and 3 nb for
the dibaryon production cross section at proton beam energies of 200 and
310 MeV, respectively.
This result agrees with the estimates of the cross
section obtained at the conditions of this experiment in the framework
of the dibaryon production model suggested in ref. \cite{prc}. 

The reactions $pd\to ppX$ and $pd\to pdX$ were studied also 
in the Research Center for Nuclear Physics at the proton energy 295 MeV 
over a mass range of 1898 to 1953 MeV \cite{tamii}. They did not observe any 
narrow structure and determined upper limits of the production cross
sections below 1$\mu$b/sr. 
These results are in contradiction to the observation at INR and
with the investigation of Tatischeff {\em et al.} \cite{tat2}.

In order to argue more convincingly that the states found are
really SNDs, an additional experimental investigation of the dibaryon
production is needed.

In ref. \cite{pi0} narrow dibaryon resonances were searched for in the
reaction $\g d\to\pi^0X$ in the photon energy region 140--300 MeV.
No significant structure was observed. Upper limits for the production
of narrow dibaryons in the range 2--5 $\mu$b averaged over the 0.8 MeV
resolution were established. However, the expected values of the SND
photoproduction cross section are bellow 5nb.
Significantly larger cross sections are expected
in charged pion photoproduction.

\begin{wrapfigure}{l}{0.45\textwidth}
\epsfxsize=0.4\textwidth  
\epsfysize=5cm            
\epsffile{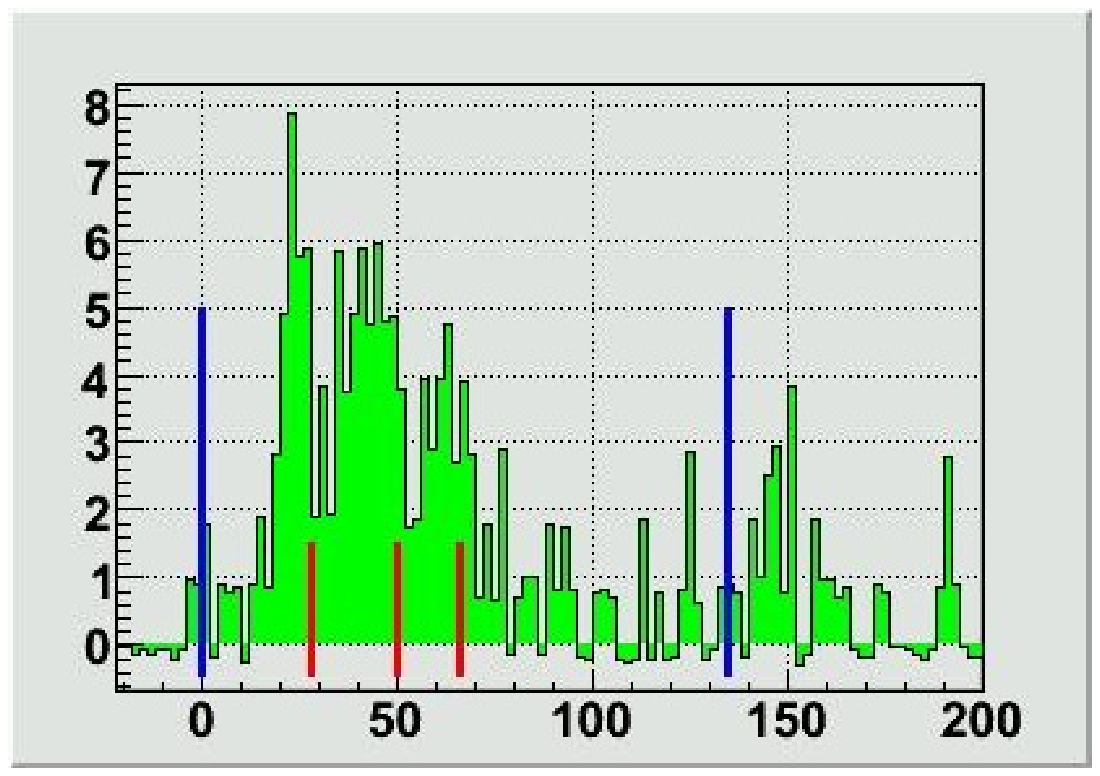}  
\caption{The missing mass $(MM(\g,\pi^0)-m_d)$ (MeV)		 
spectra of the reaction $\go$ }
\label{fig2}
\end{wrapfigure}
 In ref. \cite{alek1} a search for SNDs in the processes of pions
photoproduction by the linearly polarized photon was proposed. It was
shown a large variation between the differential cross sections of 
pions emitted in the plane parallel to the incident photon polarization 
and pions emitted in the plane perpendicular to the polarization.  

Such an experiment was performed using LEGS facility at Brookhaven National
Laboratory \cite{legs}. They
analyzed the reaction $d(\overrightarrow{\g},\pp n\g')n$ in the photon
energy range 210 -- 340 MeV. The average linear polarization of incident 
photons was $\geq 80$\%. As a result, they observed three peaks
in missing mass spectrum
when the $\pp$ was emitted in the plane parallel to the polarization of the 
incident $\overrightarrow{\g}$. The values of the mass found are very close 
to the values obtained in INR \cite{prc,epj}. 
This result supports the SND existence. However, these data were
limited by the resolution of the pion detection. 
So, they cannot be considered to be conclusive proof. 

The preliminary analysis of the missing mass distributions 
of the available data on the process $\go$, obtained at MAMI \cite{kashev}, 
demonstrates three peaks (Fig. \ref{fig2}), which good enough correspond to 
the values of the SND masses found in INR \cite{prc,epj} (red lines). 
Unfortunately, the statistic was very poor in this case.

We propose to search for the SNDs by studying the reactions $\gp$,
$\gn$, and $\go$ in the photon energy region 250--800 MeV at MAMI-C
using the tagged photon beam, the Crystal Ball/TAPS spectrometer, 
a multi-wire proportional chamber (MWPC), and PID.
The mass of SNDs will be reconstructed by the measurement of the
photon and two nucleons and will be compared with the missing mass
spectrum analysis obtained by the pion detection. 
The use of the photon beam with the energies of 300--800
MeV and the Crystal Ball/TAPS spectrometer, MWPC, and PID
will give a possibility
to suppress essentially background and to investigate a wide mass spectrum
and, particular, check the possibility of the existence of the SNDs at
$M_{pX_1}=1956$ \cite{khr} and 1982 MeV. The latter one was predicted in
\cite{epj,mass}.
This experiment can also help to understand nature of the peaks
in the $M_{X_1}$ mass spectra and clarify a possibility of existence
of exotic baryons with small masses.

\section{Supernarrow dibaryons}
\label{sec:2}

We will consider the following SNDs:
$D(T=1,J^P=1^+,S=1)$ and $D(1,1^-,0)$.
It is worth noting that the  state $(T=1, J^P=1^-)$ corresponds
to the states $^{31}P_1$ and $^{33}P_1$ in the NN channel.
The former is forbidden and
the latter is allowed for a two-nucleon state. In our work we will study
the dibaryon $D(1,1^-,0)$, a decay of which into two nucleons is forbidden
by the Pauli principle (i.e. $^{31}P_1$ state).

In the process $\g d\to \pi D$, SNDs can be produced
only if the  nucleons in the deuteron overlap sufficiently that a
6-quark state with deuteron quantum numbers can be formed. In this case, an
interaction of a photon or a meson with this state can change its
quantum numbers so that a metastable state is formed.
Therefore, the probability of the production of such dibaryons is
proportional to the probability $\eta$ of the 6-quark state existing
in the deuteron.

The magnitude of $\eta$ can be estimated 
by using the discrepancy between the theoretical and
experimental values of the deuteron magnetic moment \cite{kim,kon2}.
This method gave $\eta\le 0.03$ \cite{kon2}.
In ref. \cite{mass} the magnitude of $\eta$
was estimated by an analysis of the mass formula for SNDs using the
experimental value of the cross section of the SND production in
the process $pd\to p+pX_1$ obtained in \cite{epj,tamii}. As a result,
$\eta\approx 10^{-4}$ was obtained.

Since the energy of nucleons, produced in the decay of the SNDs under
study with $M<2m_N +m_{\pi}$, is small, it may be expected that the main
contribution to a two nucleon system should come from the
$^{31}S_0$ (virtual singlet) state. 
The results of calculations of the
decay widths of the dibaryons into $\g NN$ on the
basis of such assumptions at $\eta=0.01$ are listed in Table \ref{table1}.

\begin{table}[h]
\centering
\caption{ Decay widths of the dibaryons $D(1,1^+,1)$ and $D(1,1^-,0)$
at various dibaryon masses $M$. $\a_{t}\approx\a_{\g NN}$
\label{table1}}
\begin{tabular}{|c|c|c|c|c|c|c|c|}\hline
$M$(GeV)     & 1.90 & 1.91 & 1.93 & 1.96 & 1.98 & 2.00 & 2.013 \\ \hline
$\a_t(1,1^+)$& 0.51 & 1.57 & 6.7  & 25.6 & 48   & 81   & 109   \\
(eV)         &      &      &      &      &      &      &       \\ \hline
$\a_t(1,1^-)$& 0.13 & 0.39 & 1.67 & 6.4  & 12   & 20   & 27    \\
(eV)         &      &      &      &      &      &      &       \\ \hline
\end{tabular}
\end{table}

As a result of the SND decay through $^{31}S_0$ in
the intermediate state, the probability distribution of such decays
over the emitted photon energy $\omega$ should be characterized by a
narrow peak at the photon energy close to the maximum value.
Note that the interval of the photon energy from $\omega_m$ to
$\omega_m-1$ MeV contains about 75\% of the contribution to the width
of the decay $D(1,1^{\pm})\to \gamma NN$. This leads to a very small
relative energy of the nucleons from the SND decay 
so these nucleons will be emitted into a narrow angle cone with respect
to the direction of the SND momentum.

Such dibaryons could be produced in the processes under consideration 
if a pion is only emitted from the 6-quark state of the
deuteron. Therefore the vertexes of $d\to \pi D$ can be written as
\beq 
&&\a_{d\to\pi D(1,1^{-},0)}=\frac{g_1}{M} \sqrt{\eta}
\Phi_{\mu \nu} G^{\mu \nu} ,\\
&&\a_{d\to\pi D(1,1^+,1)}=\frac{g_2}{M}\sqrt{\eta}\varepsilon_{\mu\nu
\lambda\sigma}\Phi^{\mu\nu}G^{\lambda\sigma},\\ \nonumber
\eeq
where $\Phi_{\mu \nu}=r_{\mu}w_{\nu}-w_{\mu}r_{\nu}$,
$G_{\mu \nu}=p_{1\mu}v_{\nu}-v_{\mu}p_{1\nu}$, $w$ and $v$ are 4-vectors
of the dibaryon and deuteron polarization, respectively; and 
$r$ and $p_1$ are the dibaryon and deuteron 4-momenta.

The constants $g^2_1/4\pi$, $g^2_2/4\pi$, and $\eta$ are unknown.
However, the products of these coupling constants and $\eta$ can be
estimated from the results of work \cite{legs} where the SNDs
were searched for in the process $\overrightarrow{\g} d \to \pp D\to\g'\pp nn$.

\begin{wrapfigure}{l}{0.3\textwidth}
\centering
\epsfxsize=0.3\textwidth   
\epsffile{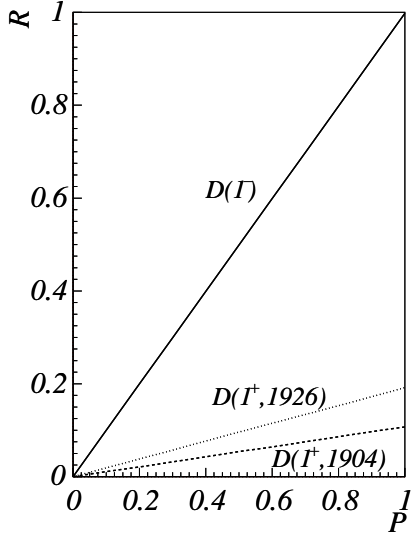}      
\caption{The asymmetry $R=(\sigma (\alpha=90^o)-\sigma (\alpha=0^o))/
(\sigma (\alpha=90^o)+\sigma (\alpha=0^o))$ 
for the process $\vec \g d\to \pi^+ D$ as a function of the 
degree of photon polarization $P$.}
\label{fig3}
\end{wrapfigure}
As a result, we have
\begin{equation}
\eta\frac{g_1^2}{4\pi}=1.4\times 10^{-4}, \qquad
\eta\frac{g_2^2}{4\pi}=3\times 10^{-4}.
\end{equation}
The cross section of the SND photoproduction by linear polarized photons
in the reaction $\overrightarrow{\g} d\to \pi D$ can be written as:
\beq
 \frac{d\sigma}{d\Omega}=A+\frac{q^2}{2}\sin^2\theta_{\pi} \, B
(1-P \,\cos 2\alpha) 
\eeq
where $\alpha$ is the angle of the photon polarization relative to
the reaction plane.
The result of the calculation of the expected asymmetry 
$R=\frac{(\sigma (\alpha=90^o)-\sigma (\alpha=0^o))}
{(\sigma (\alpha=90^o)+\sigma (\alpha=0^o))}$ 
for the process $\vec \g d\to \pi^+ D$ is shown
in Fig.~\ref{fig3} as the function of the polarization degree $P$
for SNDs $D(1,1^-)$ (for any mass), $D(1,1^+,M=1904)$, and $D(1,1^+,M=1926)$.
The largest asymmetry is expected for $D(1,1^-)$.
The asymmetry at MAMI could be about
$R\approx 0.7$ for $D(1,1^-)$.

\section{Kinematics of SND Photoproduction}

We consider SNDs with the masses $M$ in the region 1900--2000 MeV
and calculated
the total and differential cross sections
of the $D(1,1^{-})$ and $D(1,1^+)$ production in the processes of
the charged and neutral 
pion photoproduction from the deuteron.

\begin{figure}[ht]
\begin{minipage}{0.24\linewidth}       
\epsfxsize=\textwidth
\epsfysize=4.5cm
\epsffile{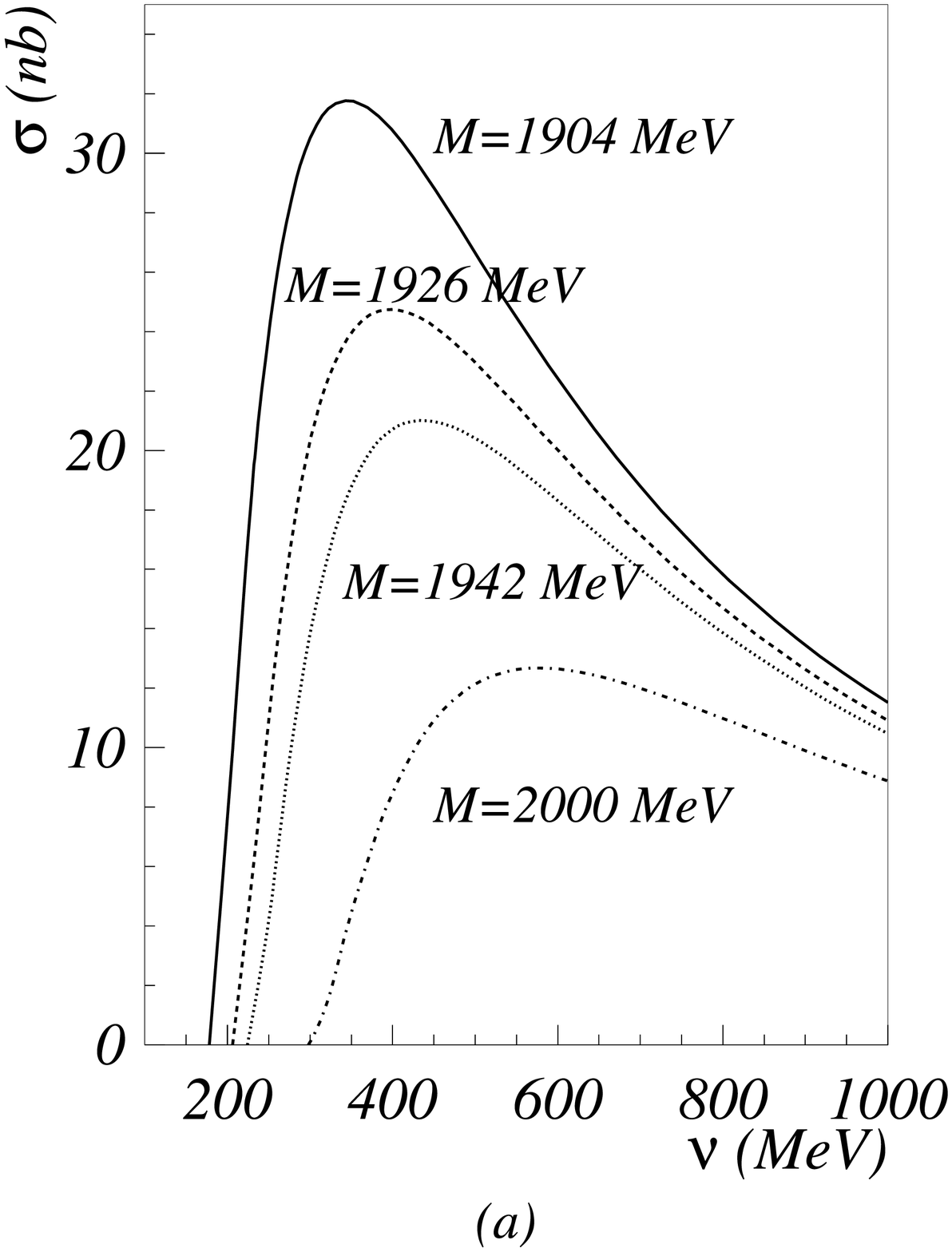}    
\end{minipage}
\begin{minipage}{0.24\textwidth}
\epsfxsize=\textwidth
\epsfysize=4.5cm
\epsffile{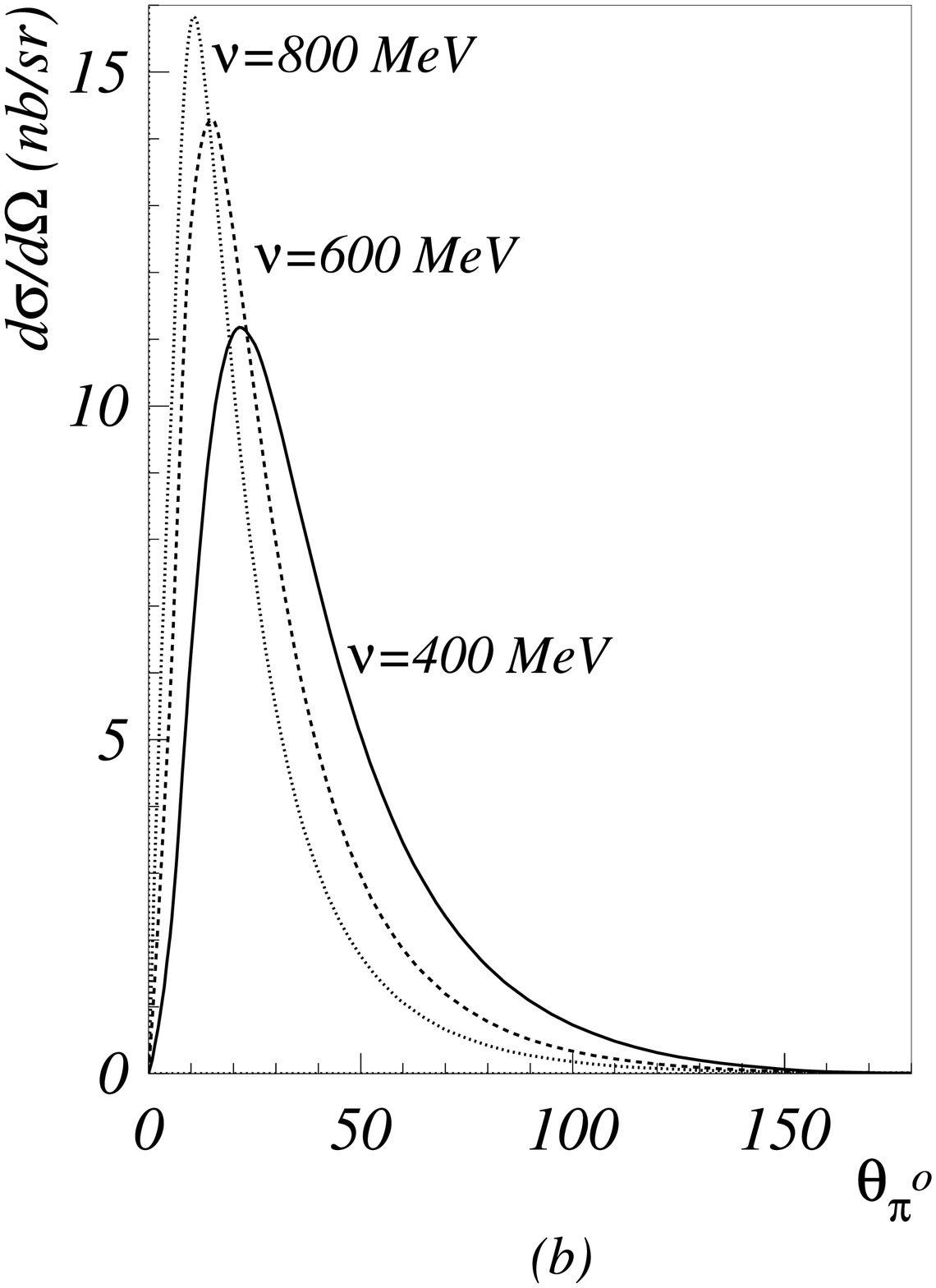}
\end{minipage}
\begin{minipage}{0.24\textwidth}
\epsfxsize=\textwidth
\epsfysize=4.5cm
\epsffile{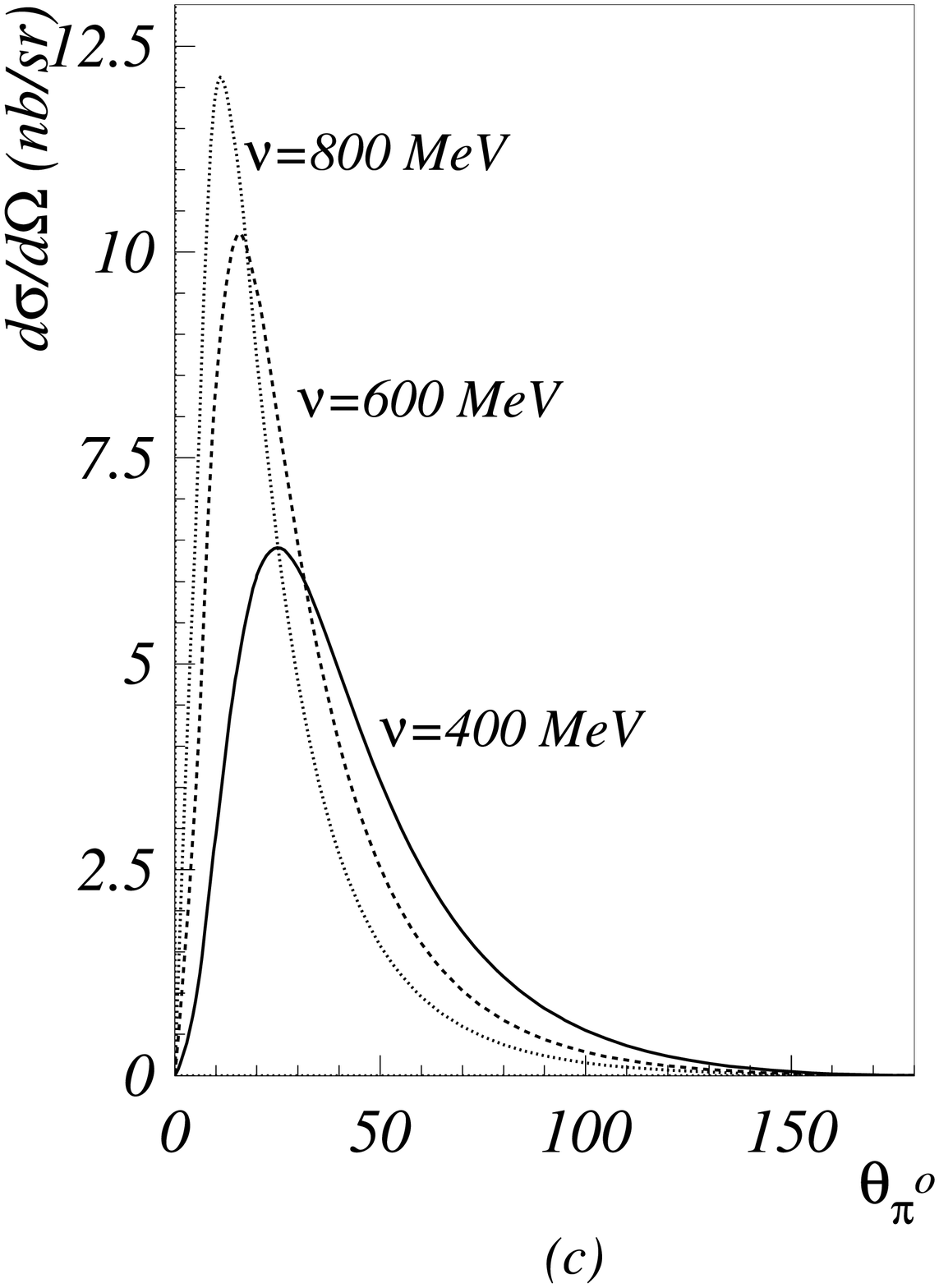}
\end{minipage}
\begin{minipage}{0.25\textwidth}
\epsfxsize=\textwidth
\epsfysize=4.5cm
\epsffile{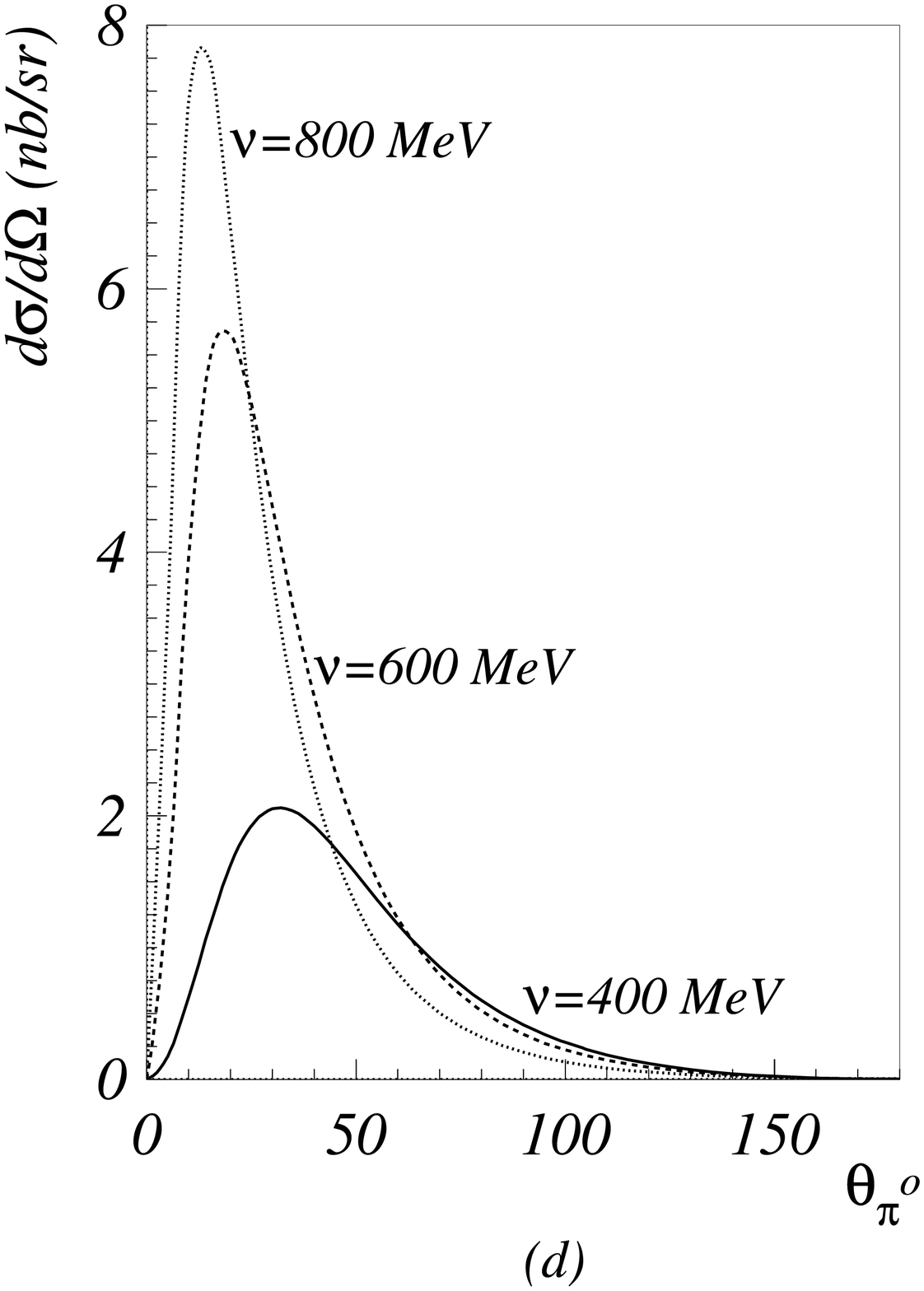}
\end{minipage}
\caption{The cross sections of the SND $D(1,1^-,0)$ production in the
reaction $\g d\to \pi^+D$; (a) --the total cross sections; (b,c,d) --
the differential cross sections for $M=1904$, 1942,, and 2000 MeV,
respectively.
\label{fig4}}
\end{figure}

\begin{figure}[ht]
\begin{minipage}{0.24\textwidth}   
\epsfxsize=\textwidth
\epsfysize=4.5cm
\epsffile{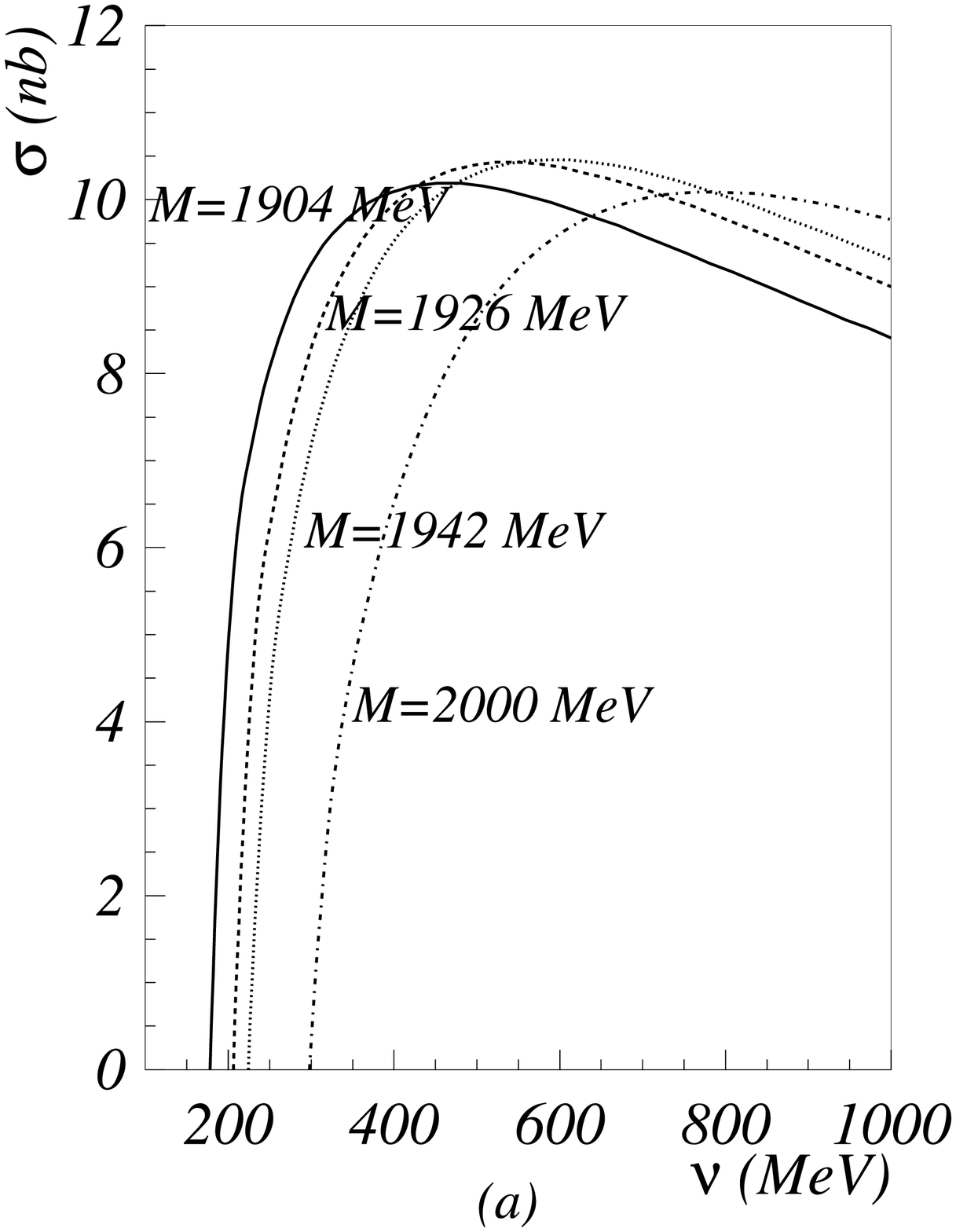}
\end{minipage}
\begin{minipage}{0.24\textwidth}
\epsfxsize=\textwidth
\epsfysize=4.5cm
\epsffile{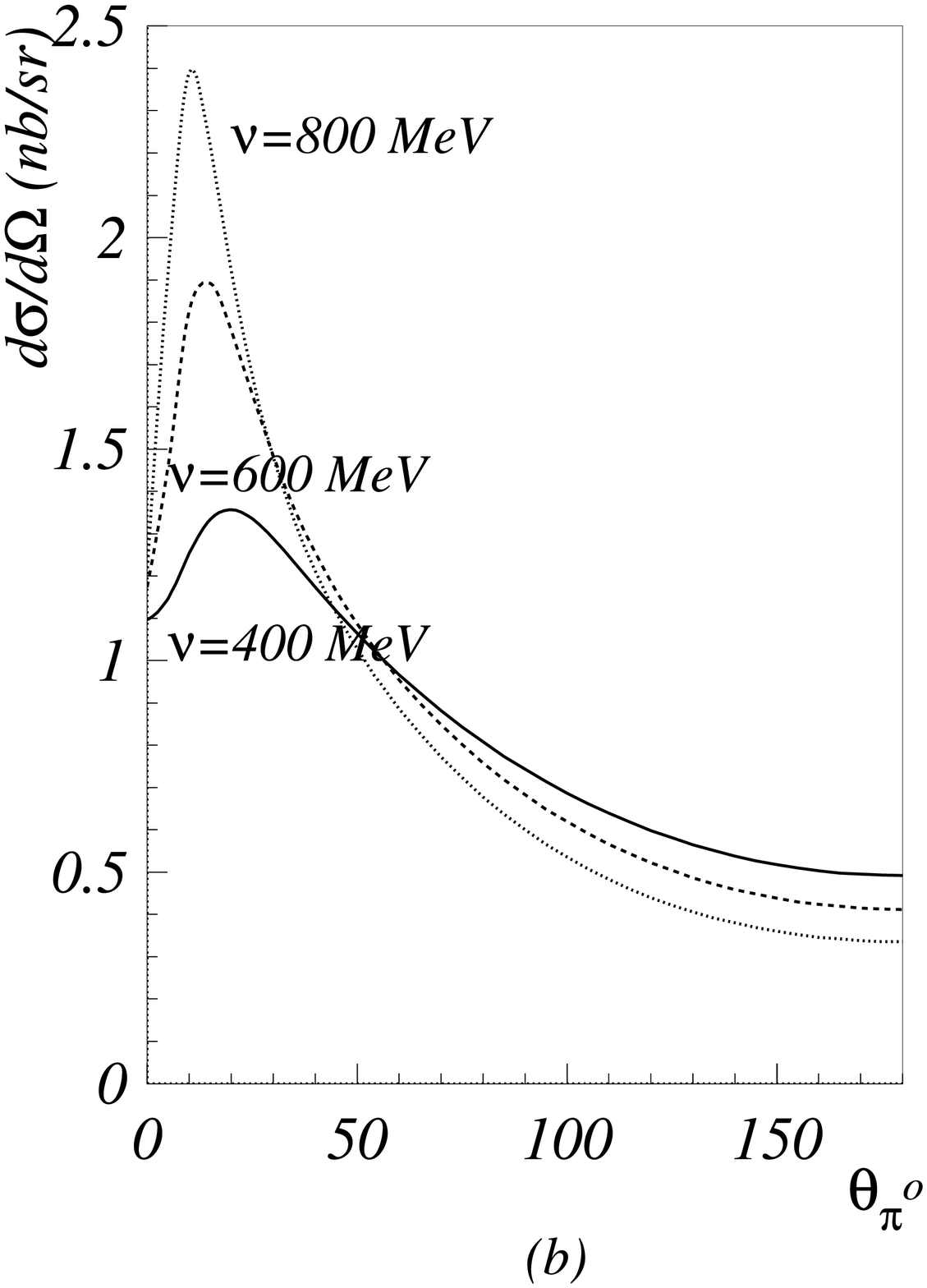}
\end{minipage}
\begin{minipage}{0.24\textwidth}
\epsfxsize=\textwidth
\epsfysize=4.5cm
\epsffile{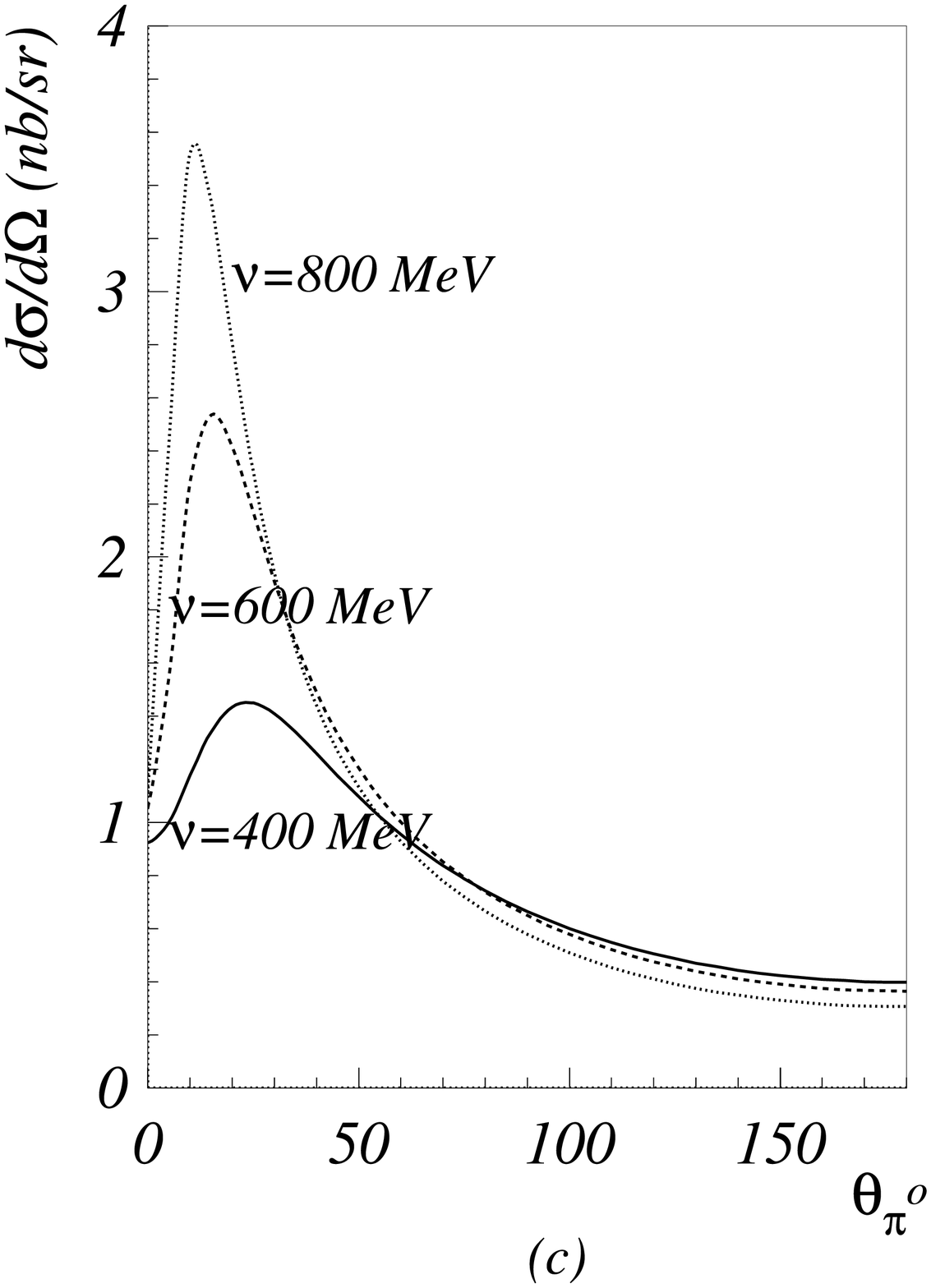}
\end{minipage}
\begin{minipage}{0.25\textwidth}
\epsfxsize=\textwidth
\epsfysize=4.5cm
\epsffile{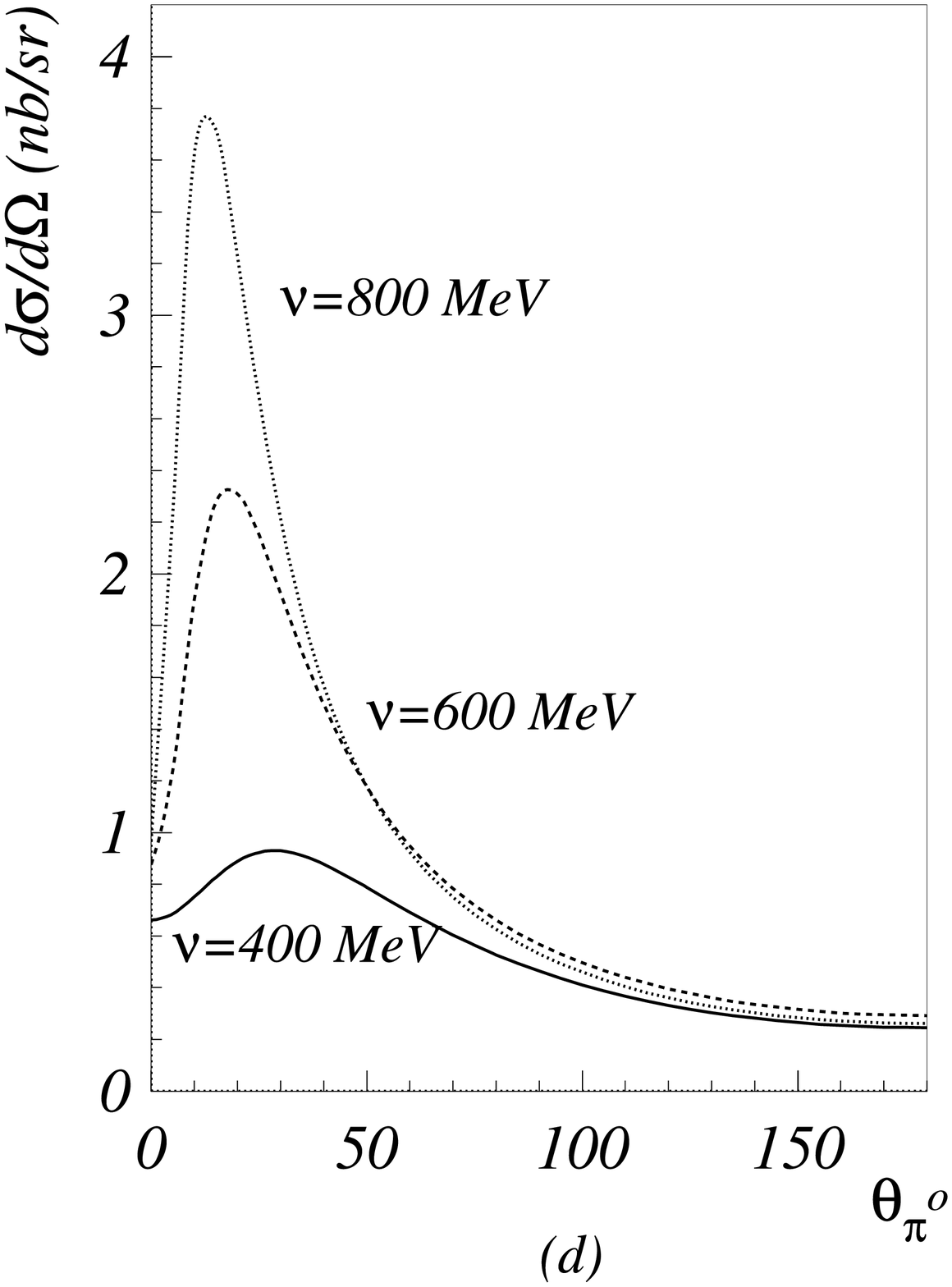}
\end{minipage}
\caption{The cross sections of the SND $D(1,1^+,1)$ production in the
reaction $\g d\to \pi^+D$.
\label{fig5}}
\end{figure}

The total cross sections for the
$D(1,1^-)$ photoproduction in the reaction with the $\pp$ meson formation
(Fig. \ref{fig4}) change from 10 nb up to 32 nb in the photon energy
interval under consideration. The main contribution to the differential
cross sections is given by pions emitted in the angular region
$0^{\circ}\div 100^{\circ}$ with maximum at $10^{\circ}-30^{\circ}$.
The total cross sections for the $D(1,1^+)$ photoproduction in this
reaction are equal to $\sim 10$ nb (Fig. \ref{fig5}).

\begin{figure}[ht]
\begin{minipage}{0.24\textwidth}
\epsfxsize=\textwidth
\epsfysize=4.5cm
\epsffile{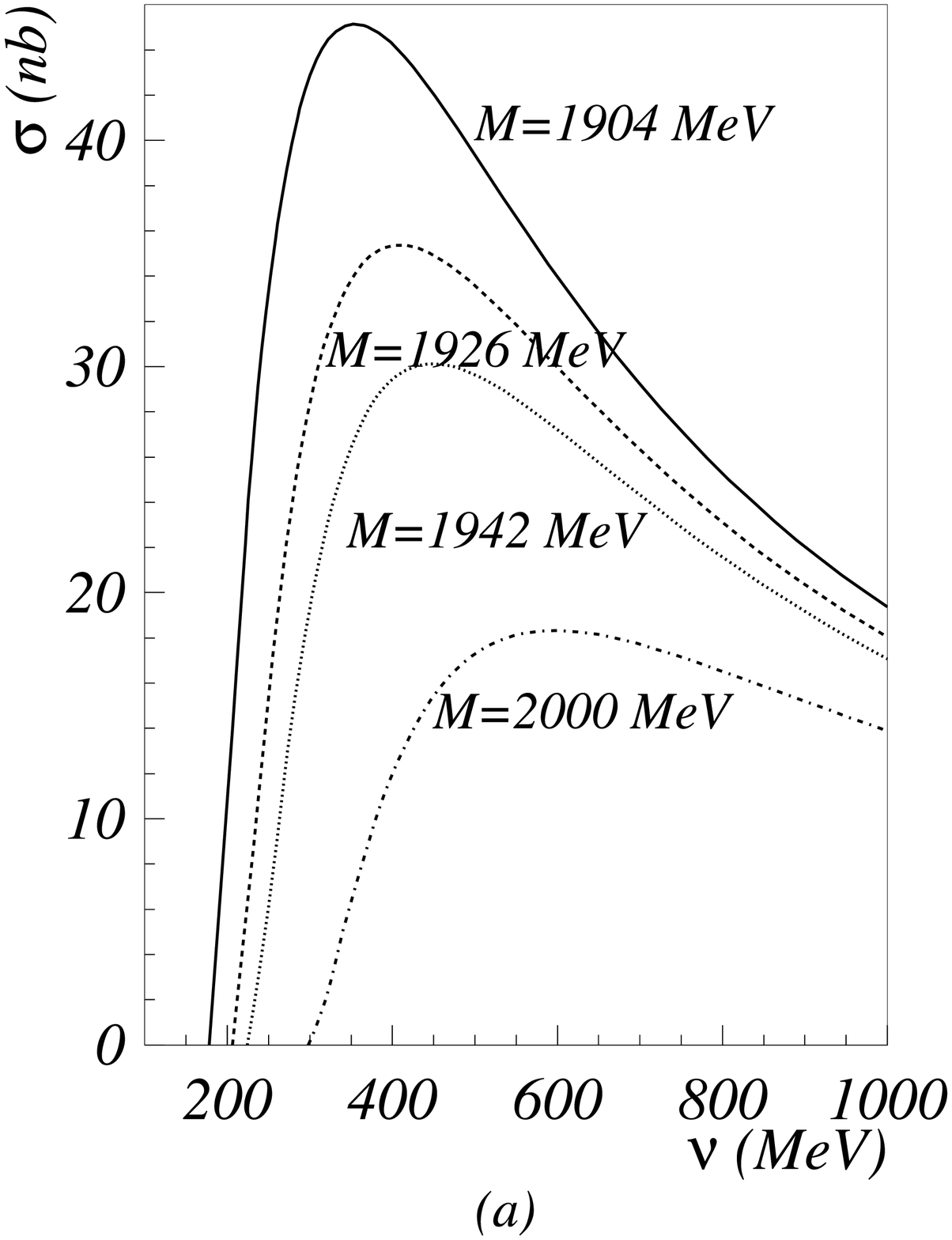}
\end{minipage}
\begin{minipage}{0.24\textwidth}
\epsfxsize=\textwidth
\epsfysize=4.5cm
\epsffile{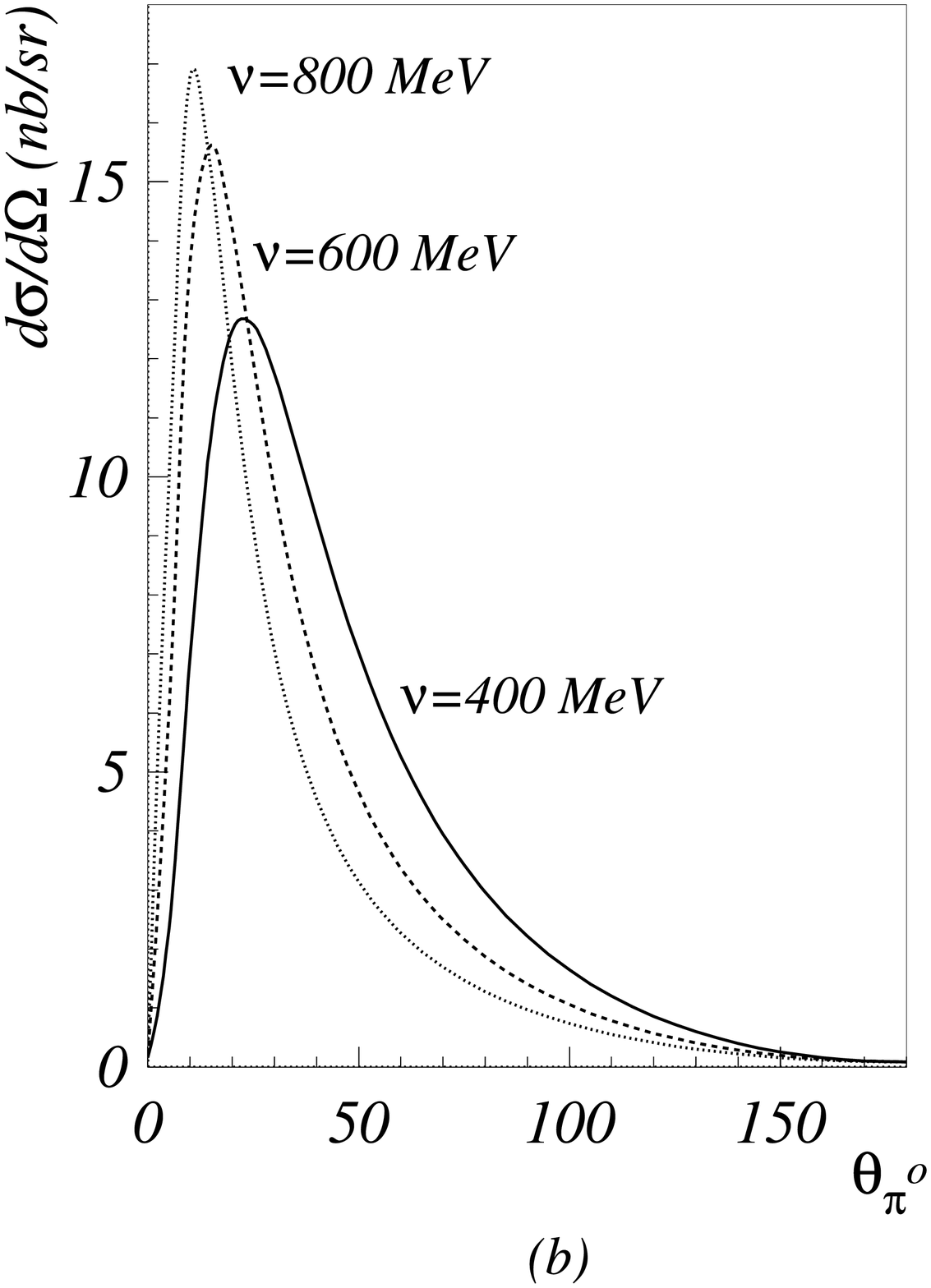}
\end{minipage}
\begin{minipage}{0.24\textwidth}
\epsfxsize=\textwidth
\epsfysize=4.5cm
\epsffile{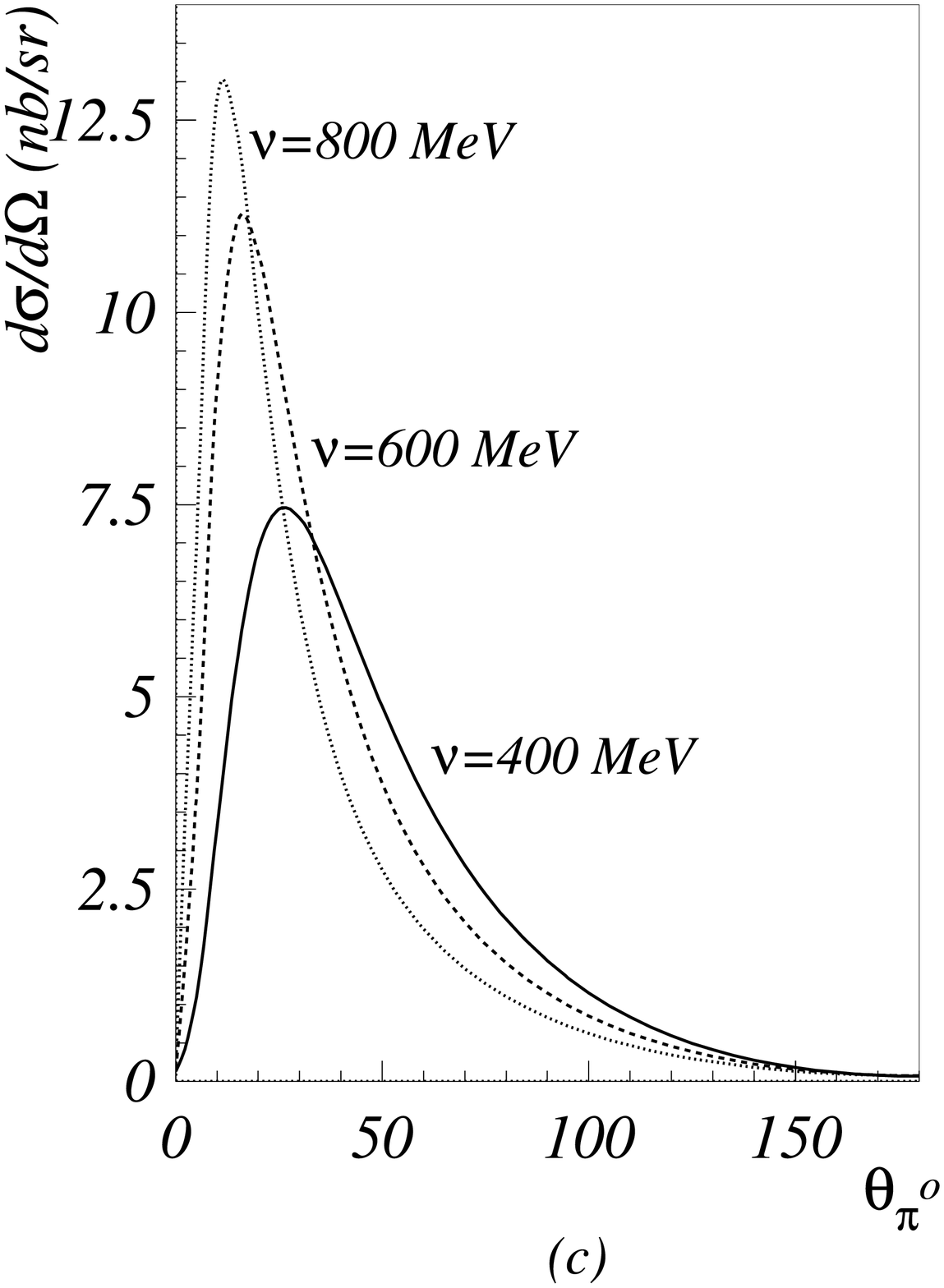}
\end{minipage}
\begin{minipage}{0.25\textwidth}
\epsfxsize=\textwidth
\epsfysize=4.5cm
\epsffile{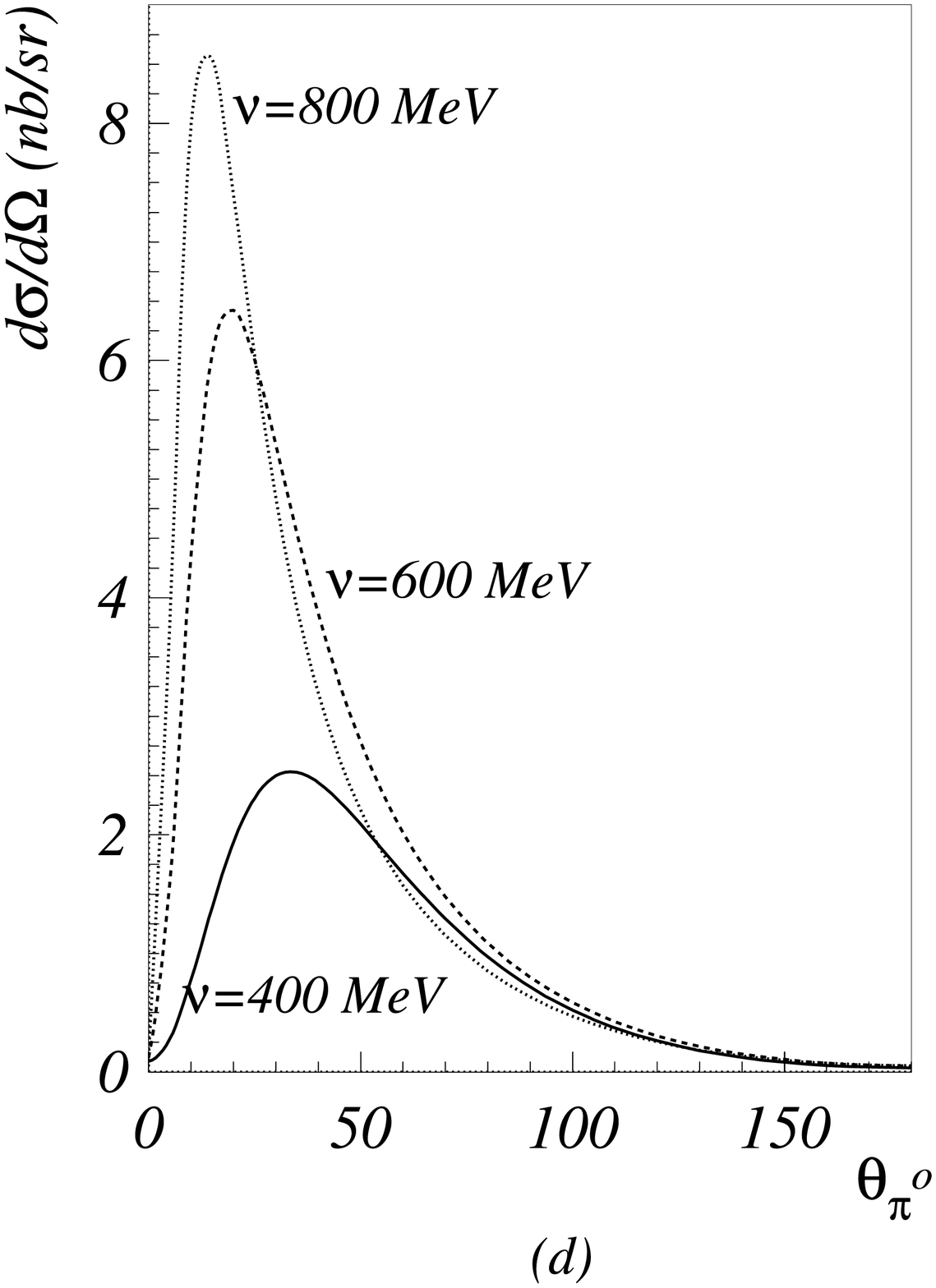}
\end{minipage}
\caption{The cross sections of the SND $D(1,1^-,0)$ production in the
reaction $\g d\to \pi^-D$. 
\label{fig6}}
\end{figure}

In the case of the $\pn$ meson photoproduction, the total cross sections
for the $D(1,1^-)$ and $D(1,1^+)$ production are greater and reach for
$M=1904$ MeV 45\,nb and 22\,nb, respectively (Fig. \ref{fig6}).

The values for the total cross sections of the $D(1,1^{\pm})$ production
in the reaction $\g d\to\po D$ are significantly less. 
However, the detection efficiency of the
$\po$ meson photoproduction by the Crystal Ball Spectrometer and TAPS
is higher than it is for the charged pions and, therefore, the good yields
are expected in this case. 

So, the comparison of the SND photoproduction cross sections in the 
reactions with the production of $\pi^+$, $\pi^-$, and $\pi^0$ mesons 
allows the quantum numbers of SND to be determined.

\begin{figure}[ht]
\begin{minipage}{0.31\textwidth}
\epsfxsize=\textwidth
\epsffile{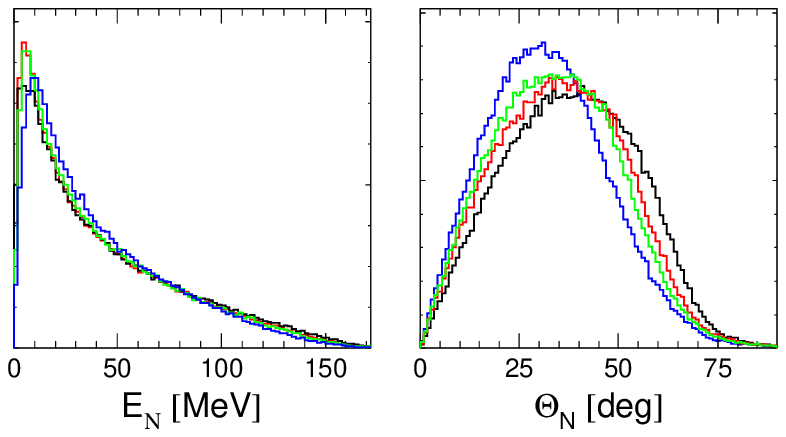}
\caption{The energy and angular distributions of
the nucleons from the decays of the dibaryons with different masses:
(black) -- $M=1900$, (red) -- 1926, (green) -- 1942,
(blue) -- 1982 MeV }
\label{fig7}
\end{minipage}
\hfill
\vspace{-2cm}
\begin{minipage}{0.31\textwidth}
\epsfxsize=\textwidth
\epsffile{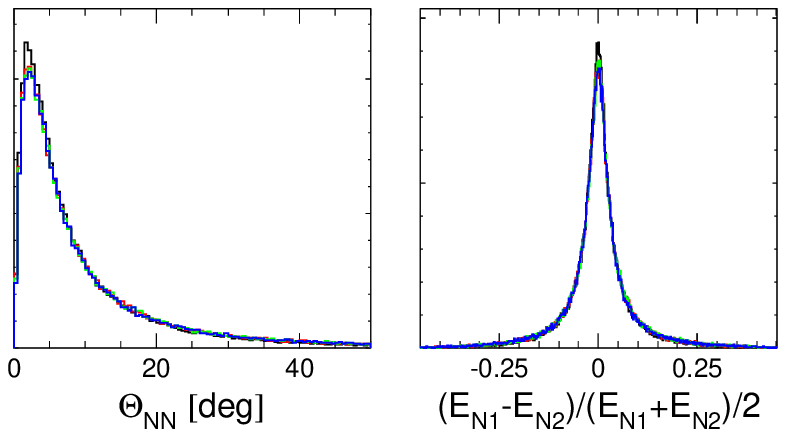}
\caption{The distributions over the angle between the final nucleons and 
over the relative
difference of the energies of these nuclons.}
\label{fig8}   
\end{minipage}
\hfill
\begin{minipage}{0.31\textwidth}
\epsfxsize=\textwidth
\epsffile{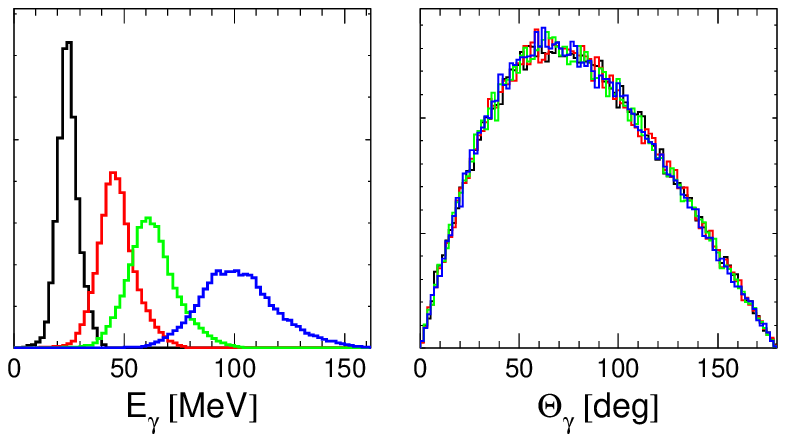}
\caption{The energy and angular distributions of
the photons from the decays of the dibaryons with different masses.}
\label{fig9}
\end{minipage}
\end{figure}

Distributions of kinematical variables for the reaction under study
are presented in Figs. \ref{fig7}-- \ref{fig9}
for the SND masses 1904, 1926, 1942, and 1982 MeV.

Fig. \ref{fig7} illustrates the expected distributions of the nucleons
over the energy and the emission angle. 
The nucleon energy is usually smaller than 100 MeV. Therefore, we will
limit ourselves by a consideration of the nucleon with such an energy.
It will permit us to suppress the background essentially. The distribution
over $\cos\theta_n$ in Fig. \ref{fig7}
shows that nucleons will be emitted primarily between
$0^{\circ}-75^{\circ}$.

Distributions over the angle between final nucleons and the relative
difference of the energies of these nucleons 
$|E_{N1}-E_{N_2}|/(E_{N1}+E_{N2})$ are presented in Fig. \ref{fig8}.
As can be seen from these figures, the angle between the two nucleons is 
below $15^{\circ}$ and the relative difference of their energies should 
be below 10\%.

Fig. \ref{fig9} demonstrates
distributions over the energy of the photon from the SND
decay and over $\cos\theta_{\g}$ for different masses of SND
(black -- $M=1904$, red -- 1926, green -- 1942, blue -- 1982 MeV).  
As seen from these the distributions for the energy of the final photon, 
the SND productions should be characterized by narrow peaks.

\section{Background}

The charged and neutral particles CB/TAPS detector combined
with the Glasgow-Mainz tagging facility is a very advantageous
instrument to search for and investigate the narrow six-quark states
in the pion photoproduction reactions. For this experimental setup
an optimal way to recognize the narrow six-quark states
is a reconstruction of the invariant mass for three particles detected:
photon and two nucleons. To suppress a background
contribution to this spectrum it is necessary to detect additionally
the pion and to determine the missing mass spectrum.

The major anticipated background reactions are
\beq\label{dph}
&& \g+d\to\po+\pp+n+n \nonumber \\
&& \g+d\to\po+\pn+p+p \\
&& \g+d\to\po+\po+p+n \nonumber 
\eeq 
\begin{equation}\label{rph}
\g+d\to\pi+\g+N+N
\end{equation}
We will limit ourselves by the
consideration of SNDs with the masses $M<2 m_N+m_{\pi}$. So, the
reactions (\ref{dph})  could be considered when only one of
the photons from the $\po$ decay is detected. But the $\po$ detection
efficiency for CB/TAPS detector is over 85\% and so few such events are
expected.

The process (\ref{rph}) has a total cross section similar to those of
the process being investigated. However, this background will be 
distributed over full missing mass region, whereas 
the processes of interest give contributions in narrow regions about 
the six-quark state masses.

In order to suppress the background more strongly, we will take into
account the following conditions:         

$\bullet$ the energy of the nucleons from the SND decay $T_N\lesssim 100$\,MeV; 

$\bullet$ the relative difference between the nucleon energies should be $\lesssim$10\%; 

$\bullet$ the angle between the nucleons should be $\Delta\theta_{NN}\approx\,15^{\circ}$; 

$\bullet$ there should be narrow peaks in the final photon energy spectrum.

\section{GEANT simulation}

To estimate expected yields for the SND formation we performed
Monte-Carlo simulation based on GEANT3 code \cite{geant}, in
which all relevant properties of the setup were taken into account.
Initial distributions for event generator included differential
cross sections of the SND production calculated according to 
the expressions obtained by us.  
The distribution of the decay photons also is
taken into account. In addition, the following beam conditions were used:

$\bullet$ Incoming electron beam energy: \qquad\qquad 855 MeV.

$\bullet$ Tagged photon energy range: \qquad\quad 250 -- 800 MeV.

$\bullet$ Maximal count rate in the tagger: \qquad $6\times 10^5$ 1/sec.

$\bullet$ Tagging efficiency:\qquad\qquad 50\%.

\begin{figure}[h]   
\begin{minipage}{0.4\textwidth}
\epsfxsize=\textwidth       
\epsffile{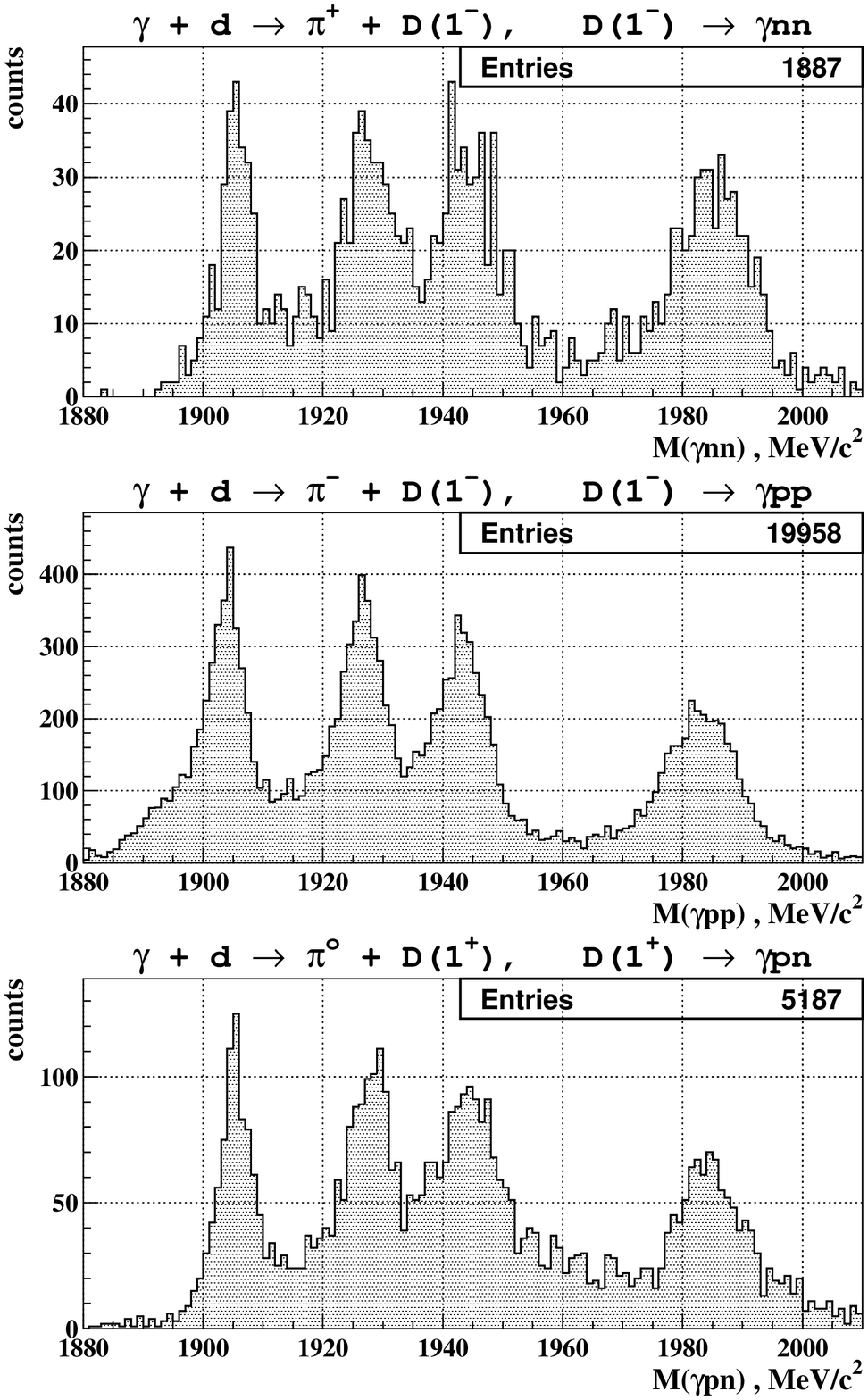}    
\caption{GEANT simulation of the SNDs with $M=1904$, 1926,
1942, and 1982 MeV production in the processes $\gp$, $\gn$,
and $\go$ for 500 hours of beam time.}
\label{fig10}
\end{minipage} 
\hfil
\begin{minipage}{0.4\textwidth}
\epsfxsize=\textwidth    
\epsffile{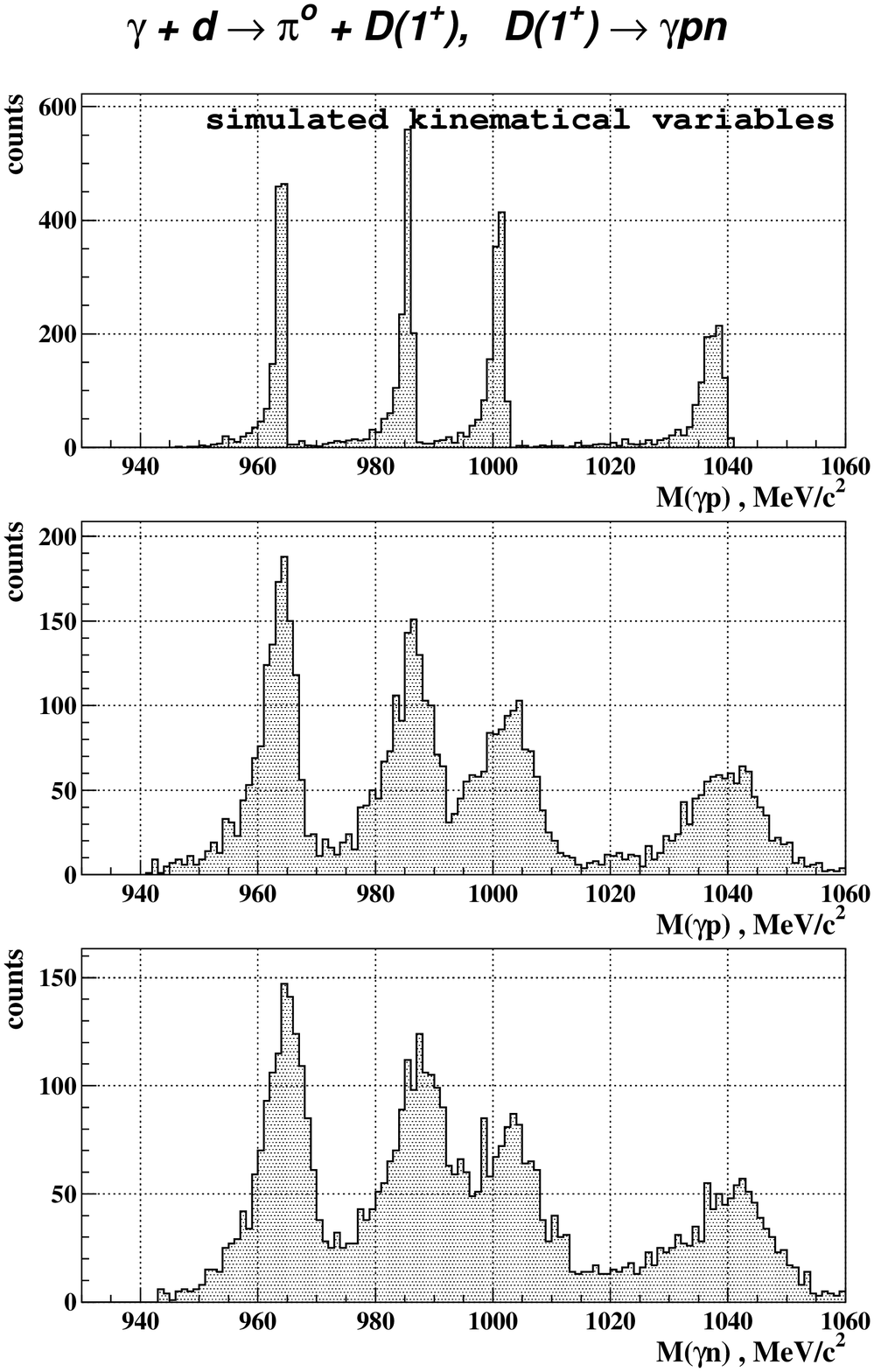}
\caption{GEANT simulation of the invariant $\g p$ and $\g n$
mass spectra for the reaction $\go$; a-- without an influence of
the detectors; b and c-- with an influence of the detectors}
\label{fig11}
\end{minipage}
\end{figure}

The results of the GEANT simulation of the invariant $\g NN$ mass
spectra for the production of the isovector SNDs with masses 1904, 1926,
1942, and 1982 MeV for the processes of the pion photoproduction
from the deuteron are presented in Fig. \ref{fig10} for 500 hours of the beam 
time. The productions of $D(1,1^-)$ in the reactions $\gp$ and $\gn$ are shown
in Fig. \ref{fig10}a and \ref{fig10}b, respectively. Fig. \ref{fig10}c shows the 
invariant $\g pn$ mass spectrum for the $D(1,1^+)$ in the process of $\pi^0$ 
meson photoproduction from the deuteron. As seen from Fig. \ref{fig10}, it is 
expected that SNDs can be easy extracted from the data with 
the number of standard deviations more than 10.

If SNDs really exist then, besides the peaks in $\g NN$ spectra, we must
also observe the corresponding peaks in $\g p$ and $\g n$ mass spectra
which connected with the dynamic of the SNDs decay. 
Such spectra, obtained
in the GEANT simulation, are represented in Fig. \ref{fig11} for the process
$\g d\to\pi^0 D(1,1^+)$, $D(1,1^+)\to \g pn$. 

The expected yields of SND as a result of the simulation of the isovector
SND production in the processes of the $\pp$, $\pn$, and $\po$-meson
photoproduction from the
deuteron at the Mainz microtron MAMI-C in the photon energy region
250--800 MeV for the beam time of 500 hours are listed in Table 2 for
the different masses of SNDs. The expected mass resolution $\sigma_M$ are
presented here too. 

\begin{table}
\centering
\begin{tabular}{|c|c|c|c|c|c|c|} \hline
$M$& \multicolumn{2}{|c|}{$\g d\to\pp D(1,1^-)$}&
\multicolumn{2}{|c|}{$\g d\to\pn D(1,1^-)$}&
\multicolumn{2}{|c|}{$\g d\to\po D(1,1^+)$} \\ \cline{2-7}
(MeV)&  &  &  &  &  &  \\
     &yields&$\sigma_M$
     &yields&$\sigma_M$
     &yields&$\sigma_M$       \\ \hline
     &   &   &    &   &   &     \\
1904 &170&4.0&2000&4.0&620&3.8 \\ \hline
     &   &   &    &   &   &     \\
1926 &210&4.6&2360&4.6&650&4.4 \\ \hline
     &   &   &    &   &   &     \\
1942 &200&5.7&2180&5.5&650&5.6 \\ \hline
     &   &   &    &   &   &     \\
1982 &240&6.9&2100&7.0&530&6.9 \\ \hline
\end{tabular}
\caption{ Expected yields of the SNDs
and the mass resolutions \mbox{($\sigma_M$ (MeV))}}
\end{table}

\section{Summary}

\noindent

$\bullet$ A search for narrow six-quark states in the reactions
$\gp$, $\gn$, and $\go$ at \mbox{MAMI-C} in the photon energy region 
250--800 MeV is proposed.

$\bullet$ The masses of SNDs will be reconstructed by the measurement
of the photon and two nucleon and will be compared with the missing mass
spectrum analysis obtained by the pion detection.

$\bullet$ Using of the Crystal Ball/TAPS spectrometer, MWPC, and PID allows 
one to detect $\g$, $p$ and $n$ with good enough accuracy and
suppress essentially the background.

$\bullet$ This experiment provides an unique opportunity to observe the SNDs 
in the mass region from 1880 up to 2000 MeV with good precision.

$\bullet$ A comparison of the results obtained for the reactions under study
will allow the quantum numbers of the SNDs ($T$, $J^P$)
to be determined.

$\bullet$ The beam time of 500 hours will allow one to determine SNDs with the
standard deviation number more than 10.

$\bullet$ Study of the $\g p$ and $\g n$ mass spectra will give
an additional information about the nature of the observed dibaryon states
and a possibility of existence of exotic baryons with small masses.

$\bullet$ Since the large beam asymmetry is expected at photoproduction of SNDs
by linear polarized photons off the deuteron, we propose first to carry
out such an experiment, which will,
in particular, allow the result obtained at LEGS \cite{legs} to be checked.

\vspace{1.5cm}
\vspace{1cm}

\vspace{1.5cm}

\newpage
\vspace{2cm}

\newpage

\end{document}